\documentclass[%
 reprint,
superscriptaddress,
preprintnumbers,
nofootinbib,
 amsmath,amssymb,
 aps,
]{revtex4-1}

\usepackage{graphicx}
\usepackage{dcolumn}
\usepackage{bm}

\usepackage{float}

\usepackage[usenames,dvipsnames]{xcolor}
\definecolor{celticsgreen}{rgb}{0, 0.478, 0.2}
\definecolor{celticsgold}{rgb}{0.58823529411, 0.21960784313, 0.1294117647}

\usepackage[colorlinks=true,citecolor=celticsgreen,linkcolor=celticsgreen,urlcolor=celticsgreen,backref=false,pdfborder={0 0 0}]{hyperref}

\begin{document}

\preprint{MIT-CTP/5759}

\title{Learning Theory Informed Priors for Bayesian Inference:\\ A Case Study with Early Dark Energy}

\author{Michael W.~Toomey}
\address{Center for Theoretical Physics, Massachusetts Institute of Technology, Cambridge, MA 02139, USA}

\author{Mikhail M.~Ivanov}
\address{Center for Theoretical Physics, Massachusetts Institute of Technology, Cambridge, MA 02139, USA}

\author{Evan McDonough}
\address{Department of Physics, University of Winnipeg, Winnipeg MB, R3B 2E9, Canada}


\begin{abstract}
    Cosmological models are often motivated and formulated in the language of particle physics, using quantities such as the axion decay constant, but tested against data using ostensibly physical quantities, such as energy density ratios, assuming uniform priors on the latter. This approach neglects priors on the model from fundamental theory, including from particle physics and string theory, such as the preference for sub-Planckian axion decay constants. We introduce a novel approach to learning theory-informed priors for Bayesian inference using normalizing flows (NF), a flexible generative machine learning technique that generates priors on model parameters when analytic expressions are unavailable or difficult to compute. As a test case, we focus on early dark energy (EDE), a model designed to address the Hubble tension. Rather than using uniform priors on the \textit{phenomenological} EDE parameters $f_{\rm EDE}$ and $z_c$, we train a NF on EDE cosmologies informed by theory expectations for axion masses and decay constants. Our method recovers known constraints in this representation while being $\sim 300,000$ times more efficient in terms of total CPU compute time. Applying our NF to \textit{Planck} and BOSS data, we obtain the first theory-informed constraints on EDE, finding $f_{\rm EDE} \lesssim 0.02$ at 95\% confidence with an $H_0$ consistent with \textit{Planck}, but in $\sim 6\sigma$ tension with SH0ES. This yields the strongest constraints on EDE to date, additionally challenging its role in resolving the Hubble tension.
\end{abstract}

\maketitle


\textit{Introduction.} --- Bayesian inference is the gold standard for data analysis in the modern era of precision cosmology, from analyses of the cosmic microwave background (CMB) \cite{Planck:2018vyg} to large-scale structure (LSS) data \cite{Troster:2019ean,Ivanov:2019hqk,DESI:2024mwx,Ivanov:2024jtl}.
A defining feature of this approach is the use of a \textit{prior} as implemented in Bayes' theorem,\footnote{See \cite{Trotta:2017wnx} for a more detailed discussion of Bayesian inference in cosmology.}
\begin{equation}
    P(\theta | d, M) = \frac{P(d | \theta, M) P(\theta|M)}{P( d | M)} 
\end{equation}
which models the posterior probability $P(\theta | d, M)$ for parameters $\theta$ of a model $M$ given data $d$. One uses data to calculate the likelihood $P(d | \theta, M)$, the probability to observe the data if $\theta$ are the true parameters, together with a \textit{chosen} prior on the distribution of model parameters,  $P(\theta | M)$, to infer the posterior.\footnote{The \textit{evidence}, $P(d | M)$, is not relevant for parameter estimation alone as its role is as a normalization factor -- to ensure that the posterior is a proper probability distribution.} Thus,  Bayes' theorem ensures that the \textit{posterior} will depend, at least in some small way, on the choice of prior. Whether this is something to worry about depends on the situation; for example,  a highly significant detection (a sharply peaked likelihood) can reasonably be expected to be robust to a change in prior.

On the other hand, in the case of a model which exhibits a weak peak in the likelihood, the prior dependence can be be important, and becomes imperative that the priors chosen are as informed as possible. This, in part, stems from the so-called ``look-elsewhere effect,'' namely that scanning across many parameters has the chance of encountering random fluctuations in the likelihood which can artificially generate a ``signal'' in the data, i.e. a mild peak in the likelihood (see discussions in the context of particle physics \cite{Lyons:2008hdc,Demortier:1099967,Gross:2010qma,Ranucci:2012ed,Bayer:2020pva,Feldman:1997qc}). Given the explicit dependence of the posterior on the choice of prior, the look-elsewhere effect can result in overestimating the significance of any peaks in the likelihood and a false detection. For this reason, one should always use an \textit{informed prior} if one is available. Of course, if there is truly a lack of any prior (beyond the most basic prior, for example, it being positive) it is typically best practice to take broad, uniform priors, or ``uninformative'' priors, on the parameters of interest. The downside is that such a prior is actually \textit{informative} and runs the risk of emphasizing any possible look-elsewhere effects that may represent an unphysical region of parameter space and would otherwise be precluded with an appropriate prior. 

This is easier said than done. One possible scenario is that information on a given parameter (or set of parameters) could exist, but for one reason or another, it is not feasible to extract that prior in an easy way. That is, the prior information is in a format that is not amenable for direct integration into a Bayesian inference pipeline. This could range from issues with the structure of any data to construct a prior that result in time consuming Markov chain Monte Carlo (MCMC) analyses. 
In cases like these, state-of-the-art machine learning algorithms, in particular normalizing flows (NF) \cite{rezende2015variational} and other generative machine learning algorithms, could be leveraged to learn a theory informed prior given their unparalleled flexibility. Indeed, just such a concept has recently been explored in the context of constructing a physically informed prior on bias parameters in the effective field theory of large scale structure~\cite{Ivanov:2024hgq}. In this work, we demonstrate via a concrete example that this technique has much promise in being applied more generally to Bayesian analyses in cosmology. 

Specifically, we apply this technique to the well studied model of Early Dark Energy (EDE) \cite{Karwal:2016vyq,Poulin:2018cxd,Agrawal:2019lmo,Lin:2019qug} which has been proposed as a solution to the Hubble tension.One of the most prominent tensions in cosmology, the Hubble tension represents a disagreement in the measured expansion rate of the Universe between early- and late-Universe probes \cite{Verde:2019ivm}. Notably, the tension between the \textit{Planck} CMB measurement and the SH0ES Cepheid-calibrated supernovae measurements of $H_0$ now sits at a significance of 5$\sigma$ \cite{Riess:2021jrx}.\footnote{However, recent results with data from the James Webb Space Telescope (JWST) suggest that there may not be a tension in the expansion rate \cite{Freedman:2024eph}. } Unless there are systematics that account for the discrepancy, one has to modify the $\Lambda$ Cold Dark Matter ($\Lambda$CDM) model to restore concordance \cite{Rigault:2014kaa,NearbySupernovaFactory:2018qkd,Addison:2017fdm,CSP:2018rag,Jones:2018vbn,Efstathiou:2020wxn,Brout:2020msh,Mortsell:2021nzg,Mortsell:2021tcx,Freedman:2021ahq,Garnavich:2022hef,Kenworthy:2022jdh,Riess:2022mme,Feeney:2017sgx,Breuval:2020trd,Javanmardi:2021viq,Wojtak:2022bct}.
While there are many proposals that try to reduce the tension, EDE has been positioned as one of the most promising proposals. EDE manages to fit CMB data with a high $H_0$ but at the expense of 
shifting other cosmological parameters, which is then disfavoured by LSS data~\cite{Hill:2020osr,Ivanov:2020ril,Goldstein:2023gnw}.

This \textit{letter} is structured as follows: We begin with a review of the underlying theory of EDE and status of constraints on the model. Next we discuss expectations from theory for the distribution of axion masses and decay constants for axion-like EDE. Following this we introduce normalizing flows and discuss how we build and train such an architecture to \emph{learn} a theory informed prior for effective EDE parameters $f_{\rm EDE}$ and $z_c$. We then review the selection of data sets used for the analyses in this work. We begin our analysis by validating our approach by comparing our results with \textit{learned} theory priors to a previous analysis where the theory parameters where directly sampled in an MCMC analysis. We then present our main results for constraints on the EDE model with the theory informed priors for a cosmological data set comprising CMB and LSS data which represents one of the most constraining results for EDE. Finally, we end with a discussion of results from our analysis and reflect on its implication for EDE and broader applicability of our approach to other cosmological analyses.

\textit{The Canonical EDE Model.} --- While the original proposal for EDE was modeled with an effective fluid \cite{Karwal:2016vyq} it is generally modeled by an axion-like scalar field \cite{Poulin:2018cxd} governed by the Klein-Gordon equation,
\begin{equation}
    \phi'' + 2 a H \phi' + V_\phi = 0,
\end{equation}
where $'$ represents a derivative with respect to conformal time.\footnote{To emphasize this particular model, this is sometimes referred to as \textit{axion-like} EDE.} The scalar starts with some initial displacement from the minimum of its potential, $\theta_i = \phi_i/f$, and is initially in slow-roll until it is released from Hubble friction when $m \sim H$. During this period the equation of state $w \sim -1$ mirrors that of dark energy. The potential used for axion-like EDE takes the form,
\begin{equation}
    V(\phi) = m^2 f^2 \left( 1 - \cos{\frac{\phi}{f}} \right)^n
    \label{EDE-Pot}
\end{equation}
where $n = 3$ is typically chosen as it is a better fit to data \cite{Poulin:2018cxd}. Altogether EDE adds three new \textit{theory} parameters: the axion mass $m$, decay constant $f$, and  initial misalignment angle $\theta_i$. Equivalently, one can also map the first two EDE \textit{theory} parameters to \textit{effective} cosmology parameters $f_{\rm EDE}$, the peak fractional energy density of EDE, and $z_c$, the location of this peak. The need for a non-standard axion-like potential is due to requirements for the energy density of EDE to redshift as radiation or faster after its peak energy injection. That is, for such a potential the average equation of state once the field begins to oscillate is
\begin{equation}
    w = \frac{n - 1}{n + 1}
\end{equation}
where $n$ can be chosen to control the decay rate of EDE. As explained in \cite{Smith:2019ihp}, the best fit for the data is $n=3$ which corresponds to $w=0.5$ after the field is released from Hubble friction.

Critically, the $n=3$ EDE model can accommodate an increased $H_0$ whilst retaining a competitive fit to \textit{Planck} CMB PR3 data relative to $\Lambda$CDM
(The fit to \textit{Planck} PR4 data is worse~\cite{doi:10.1142/S0218271824300039}).
The requisite model parameters follow from simple considerations: the timing of the EDE fixed $m\simeq H(z_{\rm eq}) \sim 10^{-27} {\rm eV}$, and the amount of EDE ($\sim 10\%$ of the energy density of the universe at $z\sim z_{\rm eq}$), fixes the decay constant $f$ to be near but sub-Planckian,  $f\sim 0.2 M_{\rm pl}$.\footnote{We denote as $M_{\rm pl}=2.435 \times 10^{18}$ GeV the reduced Planck mass.} For details see e.g. \cite{Smith:2019ihp}. Beyond the concrete model described here, there are many other realization of EDE in the literature which have similar dynamics,~Refs.~\cite{Smith:2019ihp,Agrawal:2019lmo,Lin:2019qug,Alexander:2019rsc,Sakstein:2019fmf,Das:2020wfe,Niedermann:2019olb,Niedermann:2020dwg,Niedermann:2021vgd,Ye:2020btb,Berghaus:2019cls,Freese:2021rjq,Braglia:2020bym,Sabla:2021nfy,Sabla:2022xzj,Gomez-Valent:2021cbe,Moss:2021obd,Guendelman:2022cop,Karwal:2021vpk,McDonough:2021pdg,Alexander:2022own,McDonough:2022pku,Nakagawa:2022knn,Gomez-Valent:2022bku,MohseniSadjadi:2022pfz,Kojima:2022fgo,Rudelius:2022gyu,Oikonomou:2020qah,
Tian:2021omz,Maziashvili:2021mbm,Wang:2022bmk}. In addition to increasing $H_0$,  a common feature of all EDE-like models is a shift in other $\Lambda$CDM parameters to maintain a good fit to the CMB. In particular, there are significant shifts in the physical dark matter density and $\sigma_8$, all of which can be easily understood from the perspective of CMB physics, see Refs.~\cite{Lin:2019qug,Hill:2020osr} for more details. Consequently, the added constraining power from LSS datasets results in stringent constraints on the allowed amount of EDE \cite{Hill:2020osr,Ivanov:2020ril}. For further details on EDE models and observational constraints we point the reader to some reviews on the topic \cite{doi:10.1142/S0218271824300039,Kamionkowski:2022pkx,Poulin:2023lkg}.

One point of great interest for EDE has been the choice of priors in ones MCMC analysis.
Almost all analyses of EDE have assumed broad, uniform priors on the \textit{effective} cosmology parameters for EDE and not on the \textit{theory} parameters. From a Bayesian perspective, this is perfectly fine to do if there is a lack of knowledge for parameters of interest in a model. In the case of EDE, it would seem at first glance that there is no good prior for EDE in terms of its effective parameters. However, this is not true:  we {\it do} have priors on the particle physics parameters, in particular $m$ and $f$, from considerations in quantum field theory and string theory. One may wonder why the theory priors are not used directly, as we will elaborate later assuming such priors can take significantly longer for an MCMC analysis to reach convergence.

\textit{Theory Priors on EDE Parameters.}---
The requirement of a near-Planckian decay constant is at odds with expectations from particle physics and string theory (e.g. \cite{Banks:2003sx}), which have long indicated a theoretical upper bound on the axion decay constant, stemming from original arguments against global symmetry in quantum gravity \cite{Kallosh:1995hi}, later in the context of the axion Weak Gravity Conjecture \cite{Arkani-Hamed:2006emk, Rudelius:2015xta, Brown:2015iha, Hebecker:2015zss}, and explicit string theory calculations \cite{Svrcek:2006yi, Conlon:2006tq, Cicoli:2012sz}. All of these approaches predict the existence of an instanton expansion with leading-order action $f S \simeq c M_P$ where $S$ is the instanton action and $c$ an $\mathcal{O}(1)$ constant. This predicts a tower of corrections to the EDE potential,
\begin{equation}
    V_{\rm inst}(\theta) = \sum_{q=1} ^{\infty} M_{pl}^4 e^{ - c q M_{pl}/f} \cos(q \theta).
\end{equation}
 Obviously one requires $f\ll M_{pl}$ in order for this expansion to be convergent, already in mild tension with $f\simeq 0.2 M_{pl}$. In addition to this, in order for this tower of instantons to be subdominant to the EDE potential and not ruin the finely-balanced $n=3$ structure, we require the more stringent condition $M_{pl}^4 e^{ - c  Mpl/f} \ll  m^2 f^2 $, or equivalently,  $f \log \left( \frac{M_{pl}^4}{f^2 m^2} \right) \lesssim c M_{\rm pl}$. The latter
matches the bound from the axion weak gravity conjecture given in \cite{Rudelius:2022gyu} with their $\Lambda_{UV}$ set to $M_{pl}$. The swampland distance conjecture \cite{Ooguri:2006in} applied to axions \cite{Baume:2016psm,Klaewer:2016kiy,Blumenhagen:2017cxt,Scalisi:2018eaz} similarly predicts a breakdown of effective field theory for $\Delta \phi \sim f \gtrsim M_{pl} $.

Modern string theory provides further guidance, and indeed  the statistical predictions of the string theory for axion model parameters is an area of active research \cite{Broeckel:2021dpz,Halverson:2019cmy,Mehta:2020kwu,Mehta:2021pwf,Demirtas:2021gsq}. By scanning over the state-of-art flux compactifications consistent with our observable universe, one can compute the statistical distibution of axion mass and axion decay constants. Absent any prior knowledge as to the particular compactification that is realized in nature, these distributions amount to the best known {\it theory prior} on axion parameters.  Following this procedure, string theory predicts a statistical distribution for both the axion mass and axion decay constant that are each {\it log-uniform} \cite{Broeckel:2021dpz}, namely that the mass and decay constant are uniformly distributed across many orders of magnitude. In what follows, we use this information to construct a theory-informed prior on the EDE model.

\begin{figure}
    \centering
\includegraphics[width=\linewidth]{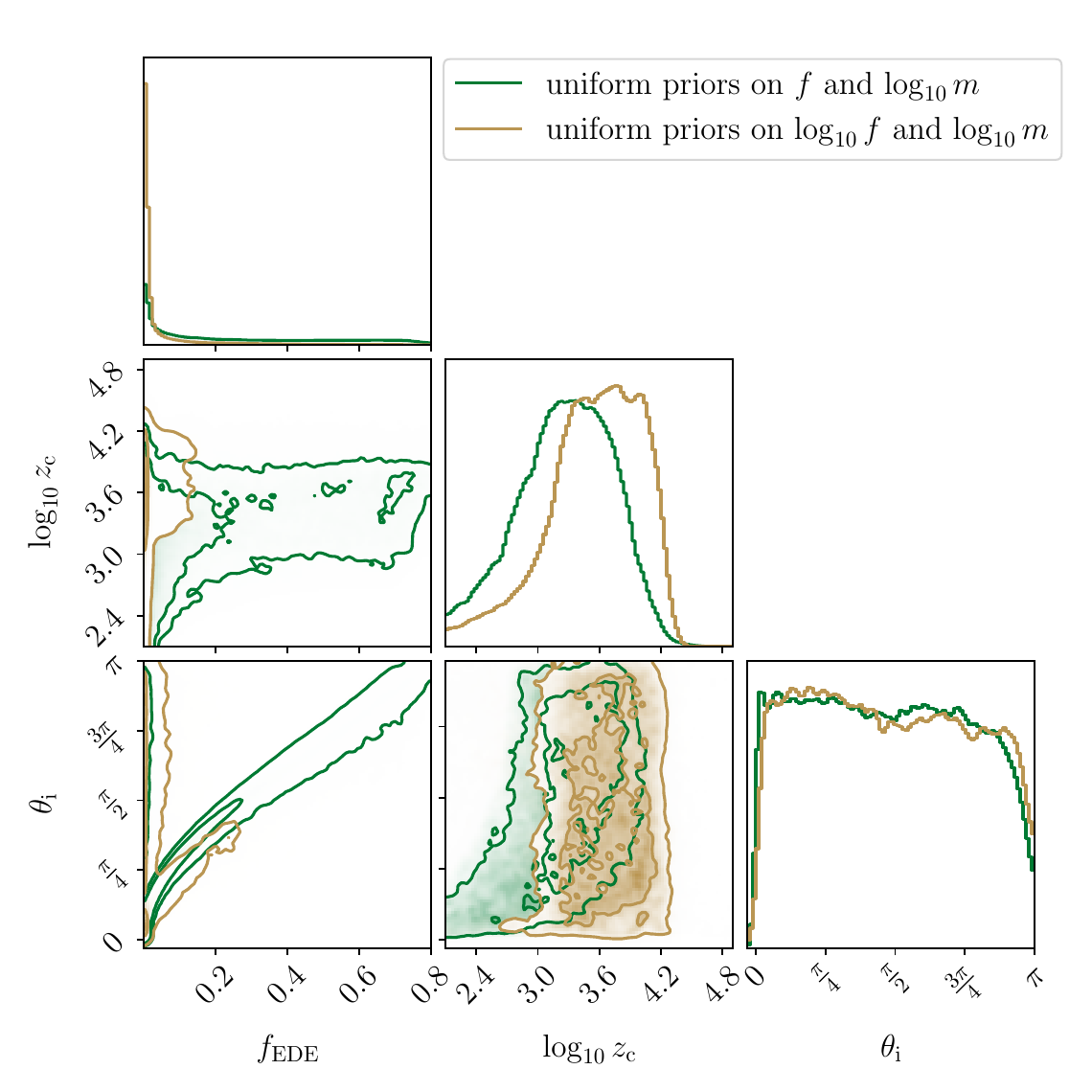}
    \caption{Effective priors from the NF learned marginal distribution for $f_{\rm EDE}$, $z_c$ and $\theta_i$ assuming theory informed priors from string theory expectation for axions (gold) and a slightly modified prior (green) that serves to validate our approach. These choice of priors correspond to Eqs.~\ref{thepri} and \ref{modthepri} respectively.}
    \label{fig:effective-priors}
\end{figure}

\textit{Normalizing Flows and Traning.}---
A form of generative machine learning, NFs~\cite{rezende2015variational} are a class of algorithms that learn a mapping $\mathcal{F}$ between a complex distribution of interest, $p(\theta)$, and simple base distribution $\pi(u)$ through a series of invertible transformations $\mathcal{F}_i$.  Once converged, the NF realizes a bijection between the target and base distributions allowing us to work directly with the simpler base distribution and seamlessly map back to $p(\theta)$. In this work we use the NFs to learn both marginalized and $\Lambda$CDM parameter conditioned distributions for the effective EDE parameters. We include further details of NFs in Supplemental Material.

To train our NF we first simulate 10,000 realizations of an EDE cosmology with \texttt{class\_ede}\footnote{\url{https://github.com/mwt5345/class_ede}} taking \textit{theory informed priors}, i.e. flat uniform priors on $\log_{10} f$, $\log_{10} m$, and $\theta_i$ 
\begin{equation}
\begin{aligned}
     \log_{10} f &\sim \mathcal{U}(10^{25},2.435 \times 10^{27}) \text{   eV} \\
     \log_{10} m &\sim \mathcal{U}(10^{-28},10^{-26}) \text{   eV} \\
     \theta_i &\sim \mathcal{U}(0.001,3.1)
\end{aligned}
\quad \text{\it theory priors}
\label{thepri}
\end{equation}
where we make a conservative choice to restrict the maximum allowed decay constant to be below $M_{\rm pl}$ (it has been suggested that the upper bound should be closer to $ f\lesssim 0.01~M_{\rm pl}$ \cite{Rudelius:2022gyu}). All assumptions which are broadly consistent with theory expectations as elaborated in the theory section.  In addition to this choice we also consider another scenario with \textit{modified theory priors} where $f$, not $\log_{10}f$, is uniformly sampled in the range,
\begin{equation}
\begin{aligned}
     f &\sim \mathcal{U}(10^{25},10^{28}) \text{   eV} \\
     \log_{10} m &\sim \mathcal{U}(10^{-28},10^{-26}) \text{   eV} \\
     \theta_i &\sim \mathcal{U}(0.1,3.1)
\end{aligned}
\quad \text{\it modified theory priors}
\label{modthepri}
\end{equation}
where we no longer impose a maximum of $M_{\rm pl}$ on $f$. We use this data set as a consistency check for our approach by comparing to a previous analysis \cite{Hill:2020osr} that sampled with these priors \textit{directly}. It should  be noted that the choice of log vs. linear prior in $f$ was made in \cite{Hill:2020osr} to precisely avoid a stronger preference for small values for $f_{\rm EDE}$; this can be seen quite clearly in Fig.~\ref{fig:effective-priors} which we discuss in detail below. For $\Lambda$CDM parameters, we match the distribution used for creation of the chains we post-process (see Supplemental Material). Altogether, our training data sets consist of 6 $\Lambda$CDM parameters $\{A_s,n_s,\tau_{\rm reio},\omega_c,\omega_b,\theta_s\}$ and 5 total EDE parameters $\{ m, f, \theta_i, f_{\rm EDE}, z_c \}$. 

In Fig.~\ref{fig:effective-priors} we display the result of our learned priors from our NFs for the marginal distribution displaying both the \textit{theory} (Eq.~\ref{thepri}) and \textit{modified theory} (Eq.~\ref{modthepri}) priors that we will use in this work. However, when we use our priors elsewhere we instead use the distribution conditioned on the $\Lambda$CDM parameters. We include further details of our NF architecture and training in the Supplemental Material. What is clear from this result (as originally pointed out in \cite{Hill:2020osr}) is that the implicit prior on effective EDE parameters is dramatically different from the choice of uniform priors. In particular, theory has a strong prior for smaller values of $f_{\rm EDE}$.

\begin{figure}[t!]
    \centering
    \includegraphics[width=1.0\linewidth]{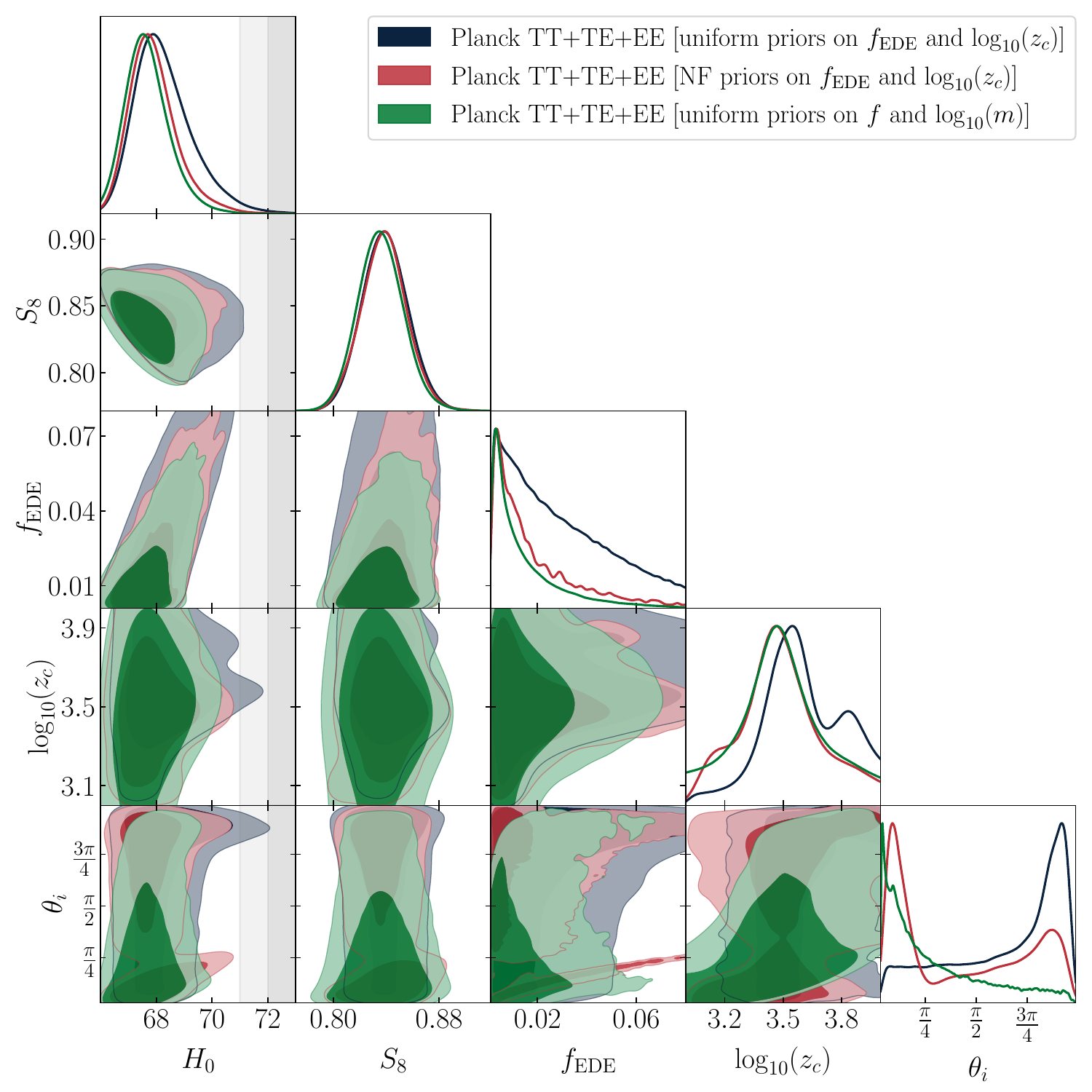}
    \caption{Parameter constraints for analysis with Planck 2018 TT+TE+EE for different choices of priors: the ``standard,'' uniform priors on $f_{\rm EDE}$ and $\log_{10} z_c$ (blue), direct, uniform priors on $f$ and $\log_{10} m$ (green) and the normalizing flow constructed \textit{modified theory priors}, i.e. Eq.~\ref{modthepri} (red), and the SH0ES $H_0$ measurement (gray band). }
    \label{fig:compare}
\end{figure}

\textit{Data.}--- We use two data sets in our analysis: \textit{Planck} 2018 low-$\ell$ and high-$\ell$ (\texttt{Plik}) temperature, polarization, and lensing power spectra with chains from \cite{Hill:2020osr}. Our second analysis consists of a combination of 
\emph{Planck} and LSS data from BOSS, eBOSS and DES with the chains from \cite{doi:10.1142/S0218271824300039}.\footnote{In the nomencloture of this paper we use \emph{Planck}+ BOSS-Full Shape P+B + BAO (2021) + eBOSS-FS-ELGs + DESY3 $S_8$ data (see~\cite{Philcox:2021kcw,Ivanov:2021zmi} for likelihoods details).}

\textit{Results}.--- As a first step we test the accuracy of the NF to correctly produce constraints from \cite{Hill:2020osr} for uniform priors on $f, \log_{10} m$ and $\theta_i$\footnote{To our knowledge this is the only time a full MCMC analysis has been conducted for EDE by sampling theory parameters directly.} by importance resampling chains from that same analysis for $f_{\rm EDE}$ and $\log_{10} z_c$ priors using the NF.\footnote{ We do so using the publicly available chains at \url{https://users.flatironinstitute.org/~chill/H20_data}} In Fig.~\ref{fig:compare} we present the updated posteriors of parameters of interest with the full result in the Supplemental Material Fig.~\ref{fig:compare-planck-full} and Table~\ref{table:params-P18-only}. Overall the results for the NF based prior are consistent with that found from sampling the particle physics parameters directly. In particular, we can see from the 1D constraints in Fig.~\ref{fig:compare} that the NF was able to closely match the constraints from directly sampling the theory prior. One may be concerned with the slight deviations in the posterior for $\theta_i$ but this is not due to our use of the NF.\footnote{The residual peak in $\theta_i$ for the NF-based posterior results from different choices in generating the MCMC chains. Specifically, a SH0ES prior was used for the generation of the \textit{effective} cosmology chains but not for the \textit{modified theory} chains in \cite{Hill:2020osr}. While this prior has been removed in post-processing, the degeneracy between high $\theta_i$ and high $H_0$ results in a relative excess of samples for large $\theta_i$ that is hard to correct for since both the effective and theory representations take uniform priors on $\theta_i$.} Overall, this is an impressive result as the mapping between constraints depends on the other $\Lambda$CDM parameters in a non-trivial way.

The remarkable success of the NF to correctly produce the constraints for the theory prior also represents a significant \textit{decrease} in the time to reach this constraint. In particular, sampling over the particle physics parameters for EDE directly can be significantly more computationally intensive than its effective representation. In particular, the constraints in \cite{Hill:2020osr} took a \textit{total} CPU time of nearly 300,000 hours or $\approx 34$ years to produce the same constraints. By contrast, analyses with effective EDE parameters take roughly a factor of $10$ less compute time. On the other hand, the theory informed constraints in this work only required an additional hour of computation with a GPU.

\begin{figure}[!t]
    \centering
    \includegraphics[width=1.0\linewidth]{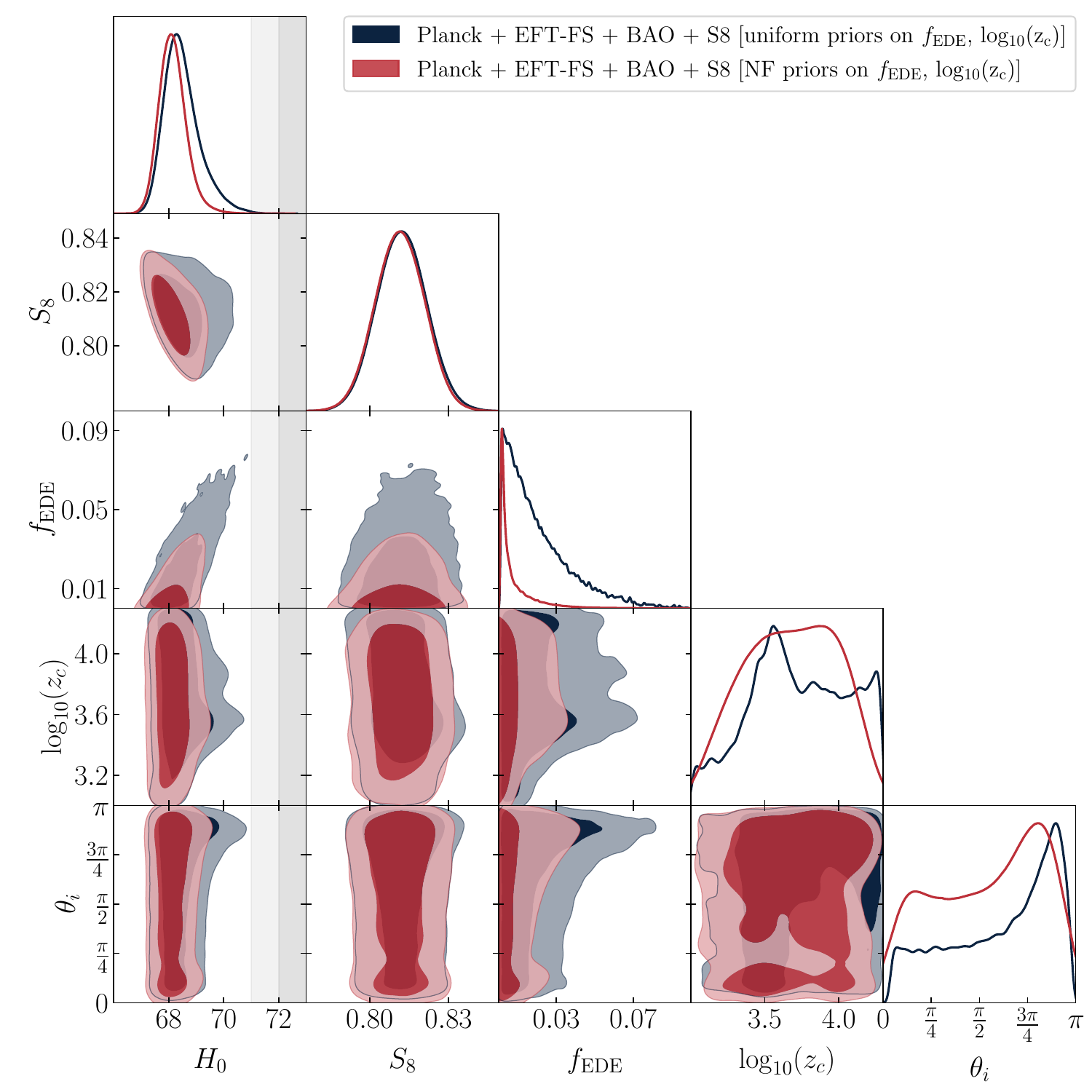}
    \caption{Constraints on cosmological parameters from analysis of {\it Planck} 2018 TT+TE+EE+low $\ell$+lensing + BOSS DR12 BAO+FS-2021 + eBOSS ELG + $S_8$ (DES-Y3) likelihood. We show the results for the standard, uniform EDE priors (blue)  and with the NF generated \textit{theory priors} (red), i.e. Eq.~\ref{thepri}, and the SH0ES $H_0$ measurement (gray band). }
    \label{fig:fs}
\end{figure}

Given the computational challenge of sampling the theory priors for EDE, post-processing MCMC chains with our NF can allow for efficient construction of posteriors with other more complicated, and time intensive, likelihoods. As a concrete example, we now explore the impact of using our NF with theory informed priors when applied to an already highly constraining data set for EDE -- \textit{Planck} combined with EFT FS with BAO from BOSS and $S_8$. In Fig.~\ref{fig:fs} we display the posterior for the \textit{effective} parameters together with $H_0$ and $S_8$ but we include the full constraints in Fig.~\ref{fig:compare-fs-full} and Table~\ref{table:params-BOSS-only} in Supplemental Material. As far as detecting an EDE component,  we find that the constraint on the allowed $f_{\rm EDE}$ is strengthened from $f_{\rm EDE} < 0.056$ with 95\% c.l. with the standard priors to $f_{\rm EDE} < 0.025$ with the NF generated theory priors (Eq.~\ref{thepri}). This strong constraint on $f_{\rm EDE}$ reflects on the other two effective EDE parameters by being unconstrained, as can be appreciated from the $\theta_i - \log_{10}z_c$ plane. On the other hand, this impacts the standard $\Lambda$CDM parameters by shifting the posterior towards the concordance model result. Of particular interest, $H_0$ is found to be $68.14^{+0.40}_{-0.52}$  which is consistent with \emph{Planck} but nearly $6\sigma$ from the SH0ES central value. This updated result represents one of the strongest constraints on the EDE scenario, comparable to recent constraints using Lyman-$\alpha$ data \cite{Goldstein:2023gnw}.

\textit{Discussion and conclusions.}--- In this \textit{letter} we have explored the application of NFs, a type of generative machine learning algorithm, as a tool to implement theory informed priors for a Bayesian inference pipeline. Using the well studied case of EDE as a testbed, we have demonstrated that NFs can correctly generate known parameter constraints by accounting for the dependence of the updated priors on other model parameters -- in this case the six standard $\Lambda$CDM parameters. Furthermore, in the particular case of EDE, our use of NFs represents a more efficient use of computational resources to achieve the same results. In this case there was a savings of over three decades of total compute time.

The case of EDE in particular showcases the importance of leveraging as much information in ones prior as possible. Indeed, for EDE the likelihood is known to be weakly peaked for a variety of data sets \cite{Herold:2022iib}. For this reason it is imperative that the prior(s) that are chosen are as informed as possible to avoid any possible ``look-elsewhere effects.'' A concern that has frequently been raised in the literature has been the potential for ``prior volume effects'' to artificially disfavor EDE but see \cite{doi:10.1142/S0218271824300039} for a critical discussion on this topic. Our results add further evidence that this is not an issue for analyses of EDE.

Regarding the status of the EDE scenario, our results with theory-informed priors with the \textit{Planck} + FS analysis strongly disfavors the viability of an axion-like EDE
resolution of the Hubble tension. Given the improved constraints in CMB only analysis for EDE with the latest \textit{Planck} likelihoods \cite{doi:10.1142/S0218271824300039}, e.g. PR4, the most up-to-date constraints would only worsen the case for EDE. 

Beyond the specific example of EDE that we have studied, it should be stressed that this technique can have a wide range of applications. Furthermore, the NF can instead be used directly in sampling for an MCMC analysis, i.e. one is not restricted to post-processing.
As a direct extension of this work, our approach can be applied to other cases where the mapping between phenomenological and physics parameters is complicated, 
e.g. inflationary
non-Gaussianity~\cite{Cabass:2022wjy,Cabass:2024wob},
and non-minimal dark 
sector models.

\textit{Acknowledgements.}--- We acknowledge helpful discussions with Stephon Alexander, Heliudson Bernardo, Cyril Creque-Sarbinowski, Carolina Cuesta-Lazaro, Sergei Gleyzer, Colin Hill, Tanvi Karwal, Siddharth Mishra-Sharma, Andrej Obuljen, and Neal Weiner. We additionally thank Fiona McCarthy for helpful comments on an early draft of this work. This material is based upon work supported by the U.S. Department of Energy, Office of Science, Office of High Energy Physics of U.S. Department of Energy under grant Contract Number  DE-SC0012567. MWT  acknowledges financial support from the Simons Foundation (Grant Number 929255). EM is supported in part by the Natural Sciences and Engineering Research Council of Canada.

\bibliography{apssamp}

\newpage 

\pagebreak
\widetext
\begin{center}
\textbf{\large Supplemental Material}
\end{center}
\setcounter{equation}{0}
\setcounter{figure}{0}
\setcounter{table}{0}
\setcounter{page}{1}
\makeatletter
\renewcommand{\theequation}{S\arabic{equation}}
\renewcommand{\thefigure}{S\arabic{figure}}
\renewcommand{\thetable}{S\Roman{table}}
\renewcommand{\bibnumfmt}[1]{[S#1]}
\renewcommand{\citenumfont}[1]{S#1}

\section{Normalizing Flows and Details of Training}

\subsection{Normalizing Flows}

A form of generative machine learning, normalizing flows \cite{rezende2015variational} are a class of algorithms that learn a mapping $\mathcal{F}$ between a complex distribution of interest, $p(\theta)$, and simple base distribution $\pi(u)$ through a series of invertible transformations $\mathcal{F}_i$. Such a transformation between two distributions can be understood in terms of the change of variables formula,
\begin{equation}
    \hat{p}(\theta)=\pi(u)\left|\operatorname{det}\left(\frac{\partial u}{\partial \theta}\right)\right|=\pi\left(\mathcal{F}_\varphi^{-1}(\theta)\right)\left|\operatorname{det} J_{\mathcal{F}_\varphi^{-1}}(\theta)\right|
\label{eq:flow}
\end{equation}
where $\left|\operatorname{det} J_{\mathcal{F}_\varphi^{-1}}(\theta)\right|$ is the Jacobian determinant of $\mathcal{F}^{-1}$ and is by construction easy to compute. Note also that $\mathcal{F}$ is taken to have a set of learnable parameters $\varphi$, in our case a neural network, that can be tuned to map from $\pi(u)$, which we will assume to be a Gaussian, and our target distribution. For a set of samples $\{\theta\}\sim p(\theta)$ the flow is trained by maximizing $\hat{p}(\theta)$ over the data, or equivalently minimizing the negative log-likelihood,
\begin{equation}
        - \log \hat{p}(\theta) = - \log \pi\left(\mathcal{F}_\varphi^{-1}(\theta)\right) - \log\left|\operatorname{det} J_{\mathcal{F}_\varphi^{-1}}(\theta)\right|
\end{equation}
where one can converge towards the optimal set of parameters $\varphi^*$ using standard machine learning techniques like stochastic gradient descent,
\begin{equation}
    \varphi^* = \underset{\varphi}{\rm arg\,max} \left<\log \hat p(\theta)\right>_{\theta\sim p(\theta)}.
\end{equation}
Once converged, the NF realizes a bijection between the target and base distributions allowing us to work directly with the simpler base distribution and seamlessly map back to $p(\theta)$. In this work we use the NFs to learn both marginalized and $\Lambda$CDM parameter conditioned distributions for the effective EDE parameters.

\subsection{Choice of Priors on $\Lambda$CDM Parameters}

Below are the priors chosen for the generation of $\Lambda$CDM parameters in addition to our two choices of prior on EDE theory parameters, Eqs.~\ref{thepri} and \ref{modthepri}. This choice matches the priors used in the generation of EDE chains from \cite{Hill:2020osr} that are used in this work.

\begin{equation}
\begin{aligned}
     \ln\left(10^{10} A_s\right) &\sim \mathcal{N}(3.05,0.001)  \\
     n_s &\sim \mathcal{N}(0.965,0.004)  \\
     100 \theta_s &\sim \mathcal{N}(1.0416,0.0004) \\
     \omega_b &\sim \mathcal{N}(0.0224,0.0001)\\
     \omega_c &\sim \mathcal{N}(0.12,0.001)\\
     \tau_{\rm reio} &\sim \mathcal{N}(0.055,0.006)
\end{aligned}
\label{lcdm-mod-priors}
\end{equation}

\subsection{Details of Training Normalizing Flows}

\begin{figure*}[!t]
    \centering
    \includegraphics[width=0.95\linewidth]{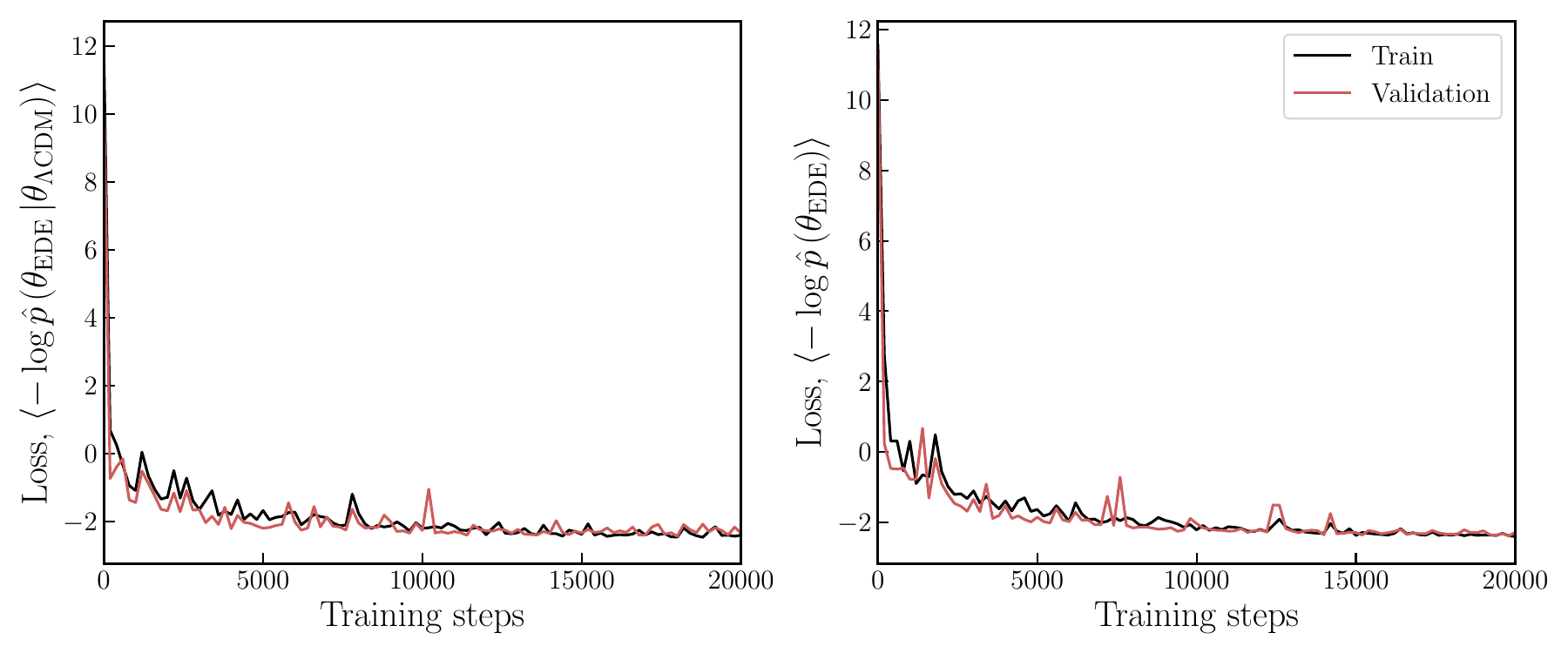}
    \caption{The training (black) and validation (red) loss for our effective EDE parameter and $\Lambda$CDM parameter conditioned  (left) and  marginal (right) normalizing flows.}
    \label{fig:loss-p-ede}
\end{figure*}

In constructing our normalizing flow we have chosen to utilize a Masked Auto-Regressive (MAF) flow \cite{papamakarios2017masked} which we use to learn the conditional probabilities for the \textit{effective} EDE parameters given the other $\Lambda$CDM parameters. We implement our architecture using the \texttt{nflows}\footnote{\url{https://github.com/bayesiains/nflows}} library and perform the training and evaluation using  \texttt{PyTorch}~\cite{paszke2019pytorch}. In our architecture we use MAF with 8 flow transformations modeled with a 2-layer masked autoregressive neural network~\cite{germain2015made} and implemented with a GELU activation function. Optimization is done with the Adam optimizer~\cite{kingma2014adam} with a learning rate of $3\times10^{-4}$ and a batch size of 1024. Our data set consists of 10,000 training samples and an additional 1,000 samples withheld for validation. Training and evaluation were conducted on an NVIDIA Tesla T4 GPU, utilizing its computational power to optimize the performance of our normalizing flow model. The results of the training for 20,000 steps are presented in Fig.~\ref{fig:loss-p-ede} which demonstrates that the log-likelihood asymptotes for the NF without overfitting. Furthermore, to ensure our data set is large enough, we have tested with 11 parameters drawn from a Gaussian distribution that 10,000 samples is sufficient to learn the underlying distribution with a normalizing flow. 

\section{Extra Tables and Plots}

\begin{table*}[h!]
Constraints from \emph{Planck} 2018 data only: TT + TE+ EE + lensing\vspace{2pt} \\
  \centering
  \begin{tabular} { l  c c c c}
\noalign{\vskip 3pt}\hline\noalign{\vskip 1.5pt}\hline\noalign{\vskip 5pt}
 \multicolumn{1}{c}{\bf } & $\Lambda$CDM &  \multicolumn{1}{c}{ Uniform $f_{\rm EDE}$ Prior} &  \multicolumn{1}{c}{ Direct (Eq.~\ref{modthepri})} &  NF Prior\\
\noalign{\vskip 3pt}\cline{2-5}\noalign{\vskip 3pt}
  Parameter &  &   &  &  \\
\hline
{\boldmath$\log{A}           $} & $3.045\pm 0.016            $ & $3.051\pm 0.017            $ & $3.046\pm 0.016            $ & $3.049\pm 0.015            $\\

{\boldmath$n_s            $} & $0.9645\pm 0.0043          $ & $0.9704^{+0.0053}_{-0.0084}$ & $0.9661^{+0.0043}_{-0.0053}$ & $0.9674^{+0.0045}_{-0.0063}          $\\

{\boldmath$100 \theta_s    $} & $1.04186\pm 0.00029        $ & $1.04163\pm 0.00034        $ & $1.04177\pm 0.00031        $ & $1.04169\pm 0.00033        $\\

{\boldmath$\omega_b        $} & $0.02235\pm 0.00015        $ & $0.02250^{+0.00018}_{-0.00022}$ & $0.02240^{+0.00015}_{-0.00018}$ & $0.02244^{+0.00017}_{-0.00021}        $\\

{\boldmath$\omega_{\rm cdm}$} & $0.1201\pm 0.0013          $ & $0.1235^{+0.0019}_{-0.0038}$ & $0.1211^{+0.0012}_{-0.0021}$ & $0.1224^{+0.0015}_{-0.0031}         $\\

{\boldmath$\tau_{\rm reio}       $} & $0.0541\pm 0.0076          $ & $0.0549\pm 0.0079          $ & $0.0543\pm 0.0077          $ & $0.0543\pm 0.0076         $\\

{\boldmath$f_{\rm EDE}    $} & -- & $< 0.087$ & $<0.041$ & $<0.048   $\\

{\boldmath$\log_{10}z_c   $} & -- & $3.67^{+0.24}_{-0.28}      $ & $<3.89              $ & $<4.00      $\\

{\boldmath$\theta_i       $} & -- & $ > 0.36      $ & $<2.31         $ & $< 2.68     $\\

{\boldmath$H_0            $} & $67.30\pm 0.60             $ & $68.30^{+0.73}_{-1.2}      $ & $67.58^{+0.59}_{-0.77}     $ & $ 67.90^{+0.62}_{-0.97}               $\\

{\boldmath$\Omega_m        $} & $0.3161\pm 0.0083          $ & $0.3144\pm 0.0086          $ & $0.3158\pm 0.0083          $ & $0.3157\pm 0.0083          $\\

\boldmath$S_8                       $ & $0.833\pm 0.016            $ & $0.839\pm 0.017            $ & $0.835\pm 0.016            $ & $0.838\pm 0.017            $\\
\hline
\end{tabular}
\caption{The mean $\pm1\sigma$ constraints on the cosmological parameters in $\Lambda$CDM and in the EDE scenario with $n=3$ for different choices of priors, as inferred from \emph{Planck} 2018 primary CMB data only (TT+TE+EE).  Upper and lower limits are quoted at 95\% CL.}
  \label{table:params-P18-only}
\end{table*}

\begin{figure}[h!]
    \centering
    \includegraphics[width=\linewidth]{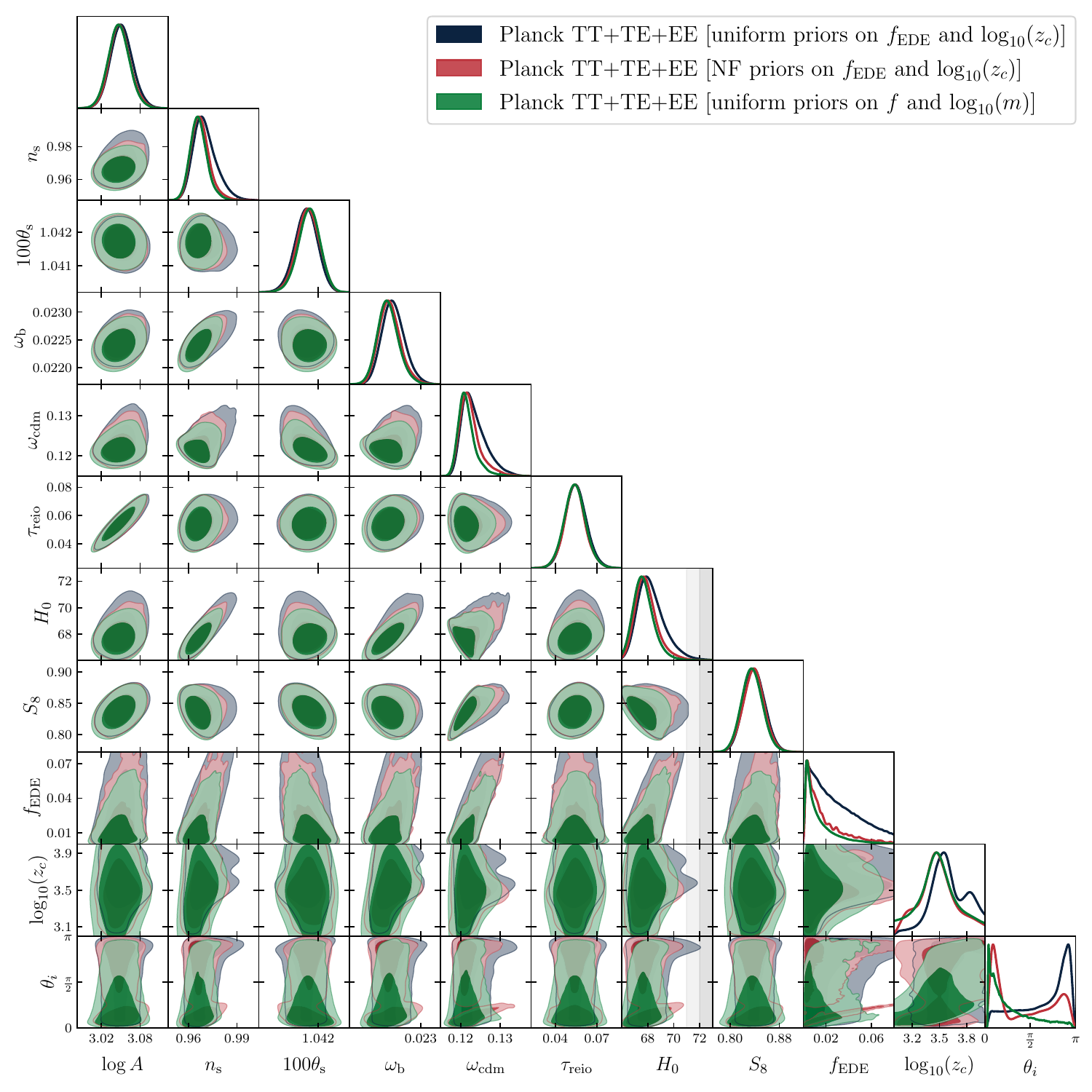}
    \caption{Parameter constraints for analysis with Planck 2018 TT+TE+EE for different choices of priors: the ``standard,'' uniform priors on $f_{\rm EDE}$ and $\log_{10} z_c$ (blue), direct, uniform priors on $f$ and $\log_{10} m$ (green) and the normalizing flow constructed \textit{modified theory priors}, i.e. Eq.~\ref{modthepri} (red), and the SH0ES $H_0$ measurement (gray band). }
    \label{fig:compare-planck-full}
\end{figure}


\begin{table*}[h!]
Constraints from {\it Planck} + EFT-FS + BAO + $S_8$  \vspace{2pt} \\
  \centering  
  \begin{tabular} { l  c c c}
\noalign{\vskip 3pt}\hline\noalign{\vskip 1.5pt}\hline\noalign{\vskip 5pt}
 \multicolumn{1}{c}{\bf } & $\Lambda$CDM &  \multicolumn{1}{c}{ Uniform $f_{\rm EDE}$ Prior} &  \multicolumn{1}{c}{Theory Prior (NF)} \\
\noalign{\vskip 3pt}\cline{2-4}\noalign{\vskip 3pt}
  Parameter &  &   &   \\
\hline
{\boldmath$\log{A}           $} & $3.036_{-0.014}^{+0.014}  $         & $3.033\pm 0.014            $ & $3.031\pm 0.014            $\\

{\boldmath$n_s            $} & $0.9642_{-0.0037}^{+0.0038} $          & $0.9692^{+0.0043}_{-0.0057}$ & $0.9675\pm 0.0042          $\\

{\boldmath$100 \theta_s    $} & $1.042_{-0.0003}^{+0.00027}      $ & $1.04175^{+0.00035}_{-0.00030}$ & $1.04186^{+0.00032}_{-0.00028}$\\

{\boldmath$\omega_b        $} & $0.02233_{-0.00014}^{+0.00013}        $ & $0.02254^{+0.00016}_{-0.00018}   $ & $0.02248^{+0.00014}_{-0.00015}   $\\

{\boldmath$\omega_{\rm cdm}$} & $0.12_{-0.00097}^{+0.00094}          $ & $0.12037^{+0.00097}_{-0.0020}$ & $0.11932^{+0.00086}_{-0.0012}$\\

{\boldmath$\tau_{\rm reio}       $} & $0.05102_{-0.0065}^{+0.0071}         $  & $0.0499\pm 0.0071          $ & $0.0498^{+0.0072}_{-0.0065}$\\

{\boldmath$f_{\rm EDE}    $} & -- & $< 0.056$ & $< 0.025$\\

{\boldmath$\log_{10}z_c   $} & -- & $3.73^{+0.55}_{-0.32}      $ & $3.70^{+0.33}_{-0.29}      $\\

{\boldmath$\theta_i       $} & -- & $1.93^{+1.1}_{-0.44}       $ & $1.68\pm 0.90              $\\

{\boldmath$H_0            $} & $67.34_{-0.42}^{+0.44}            $ & $68.50^{+0.44}_{-0.76}     $ & $68.14^{+0.40}_{-0.52}     $\\

{\boldmath$\Omega_m        $} & $0.3153_{-0.0059}^{+0.0059}          $ & $0.3060\pm 0.0054          $ & $0.3069\pm 0.0053          $\\

\boldmath$S_8                       $ & $0.8281\pm 0.011            $ & $0.8123\pm 0.0094          $ & $0.8115\pm 0.0092          $\\

\hline
\end{tabular}
\caption{The mean $\pm1\sigma$ constraints on the cosmological parameters in $\Lambda$CDM and in the EDE scenario with $n=3$ for different choices of priors, as inferred from \emph{Planck}+ BOSS-Full Shape P+B + BAO (2021) + eBOSS-FS-ELGs + DESY3 $S_8$ data (see~\cite{Philcox:2021kcw,Ivanov:2021zmi} for likelihoods details).  Upper and lower limits are quoted at 95\% CL.}
  \label{table:params-BOSS-only}
\end{table*}

\begin{figure}[b!]
    \centering
    \includegraphics[width=\linewidth]{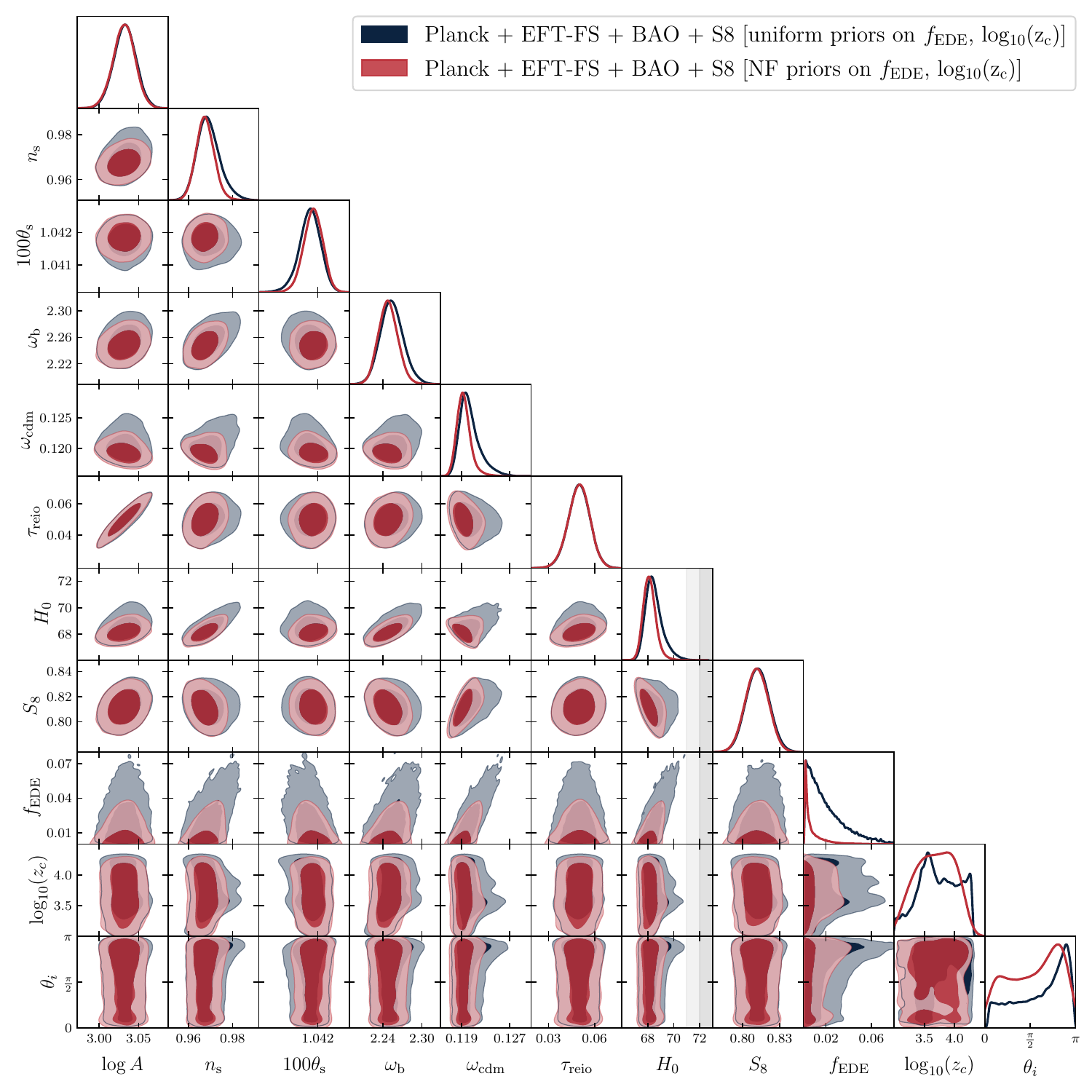}
    \caption{Constraints on cosmological parameters from analysis of {\it Planck} 2018 TT+TE+EE+low $\ell$+lensing + BOSS DR12 BAO+FS-2021 + eBOSS ELG + $S_8$ (DES-Y3) likelihood. We show the results for the standard, uniform EDE priors (blue)  and with the NF generated \textit{theory priors} (red), i.e. Eq.~\ref{thepri}, and the SH0ES $H_0$ measurement (gray band).}
    \label{fig:compare-fs-full}
\end{figure}

\end{document}